\documentclass[letterpaper]{article} 
\usepackage{test}  

\usepackage{times}  
\usepackage{helvet}  
\usepackage{courier}  
\usepackage[hyphens]{url} 
\usepackage{graphicx} 
\usepackage{natbib}  
\usepackage{caption} 
\usepackage{algorithm}
\usepackage{algpseudocode}
\usepackage{amsmath}
\usepackage{amsfonts}

\usepackage{booktabs}
\usepackage{siunitx} 

\usepackage{enumitem}

\usepackage{multirow}       
\usepackage{threeparttable} 
\usepackage{makecell}       
\usepackage{graphicx}      
\usepackage{array} 
\usepackage{caption}
\usepackage{subcaption}
\usepackage[table]{xcolor}

\usepackage{newfloat}
\usepackage{listings}

\title{ViTAD: Timing Violation-Aware Debugging of RTL Code using Large Language Models
}

\author {
    Wenhao Lv\textsuperscript{\rm 1},
    Yingjie Xia\textsuperscript{\rm 1},
    Xiyuan Chen\textsuperscript{\rm 1},
    Li Kuang\textsuperscript{\rm 2},
}
\affiliations {
    \textsuperscript{\rm 1}Micro-Electronics Research Institute, Hangzhou Dianzi University\\
    \textsuperscript{\rm 2}School of Computer Science and Engineering, Central South University\\
}

\usepackage{bibentry}

\begin{document}

\maketitle

\begin{abstract}

In modern Very Large Scale Integrated (VLSI) circuit design flow, the Register-Transfer Level (RTL) stage presents a critical opportunity for timing optimization. Addressing timing violations at this early stage is essential, as modern systems demand higher speeds, where even minor timing violations can lead to functional failures or system crashes. However, traditional timing optimization heavily relies on manual expertise, requiring engineers to iteratively analyze timing reports and debug. To automate this process, this paper proposes ViTAD, a method that efficiently analyzes the root causes of timing violations and dynamically generates targeted repair strategies. Specifically, we first parse Verilog code and timing reports to construct a Signal Timing Dependency Graph (STDG). Based on the STDG, we perform violation path analysis and use large language models (LLMs) to infer the root causes of violations. Finally, by analyzing the causes of violations, we selectively retrieve relevant debugging knowledge from a domain-specific knowledge base to generate customized repair solutions. To evaluate the effectiveness of our method, we construct a timing violation dataset based on real-world open-source projects. This dataset contains 54 cases of violations. Experimental results show that our method achieves a 73.68\% success rate in repairing timing violations, while the baseline using only LLM is 54.38\%. Our method improves the success rate by 19.30\%.

\end{abstract}

\section{Introduction}

In modern Very Large Scale Integrated (VLSI) design flow, timing optimization at the Register-Transfer Level (RTL) stage is critical to design success~\cite{fang2023masterrtl}. However, as circuit complexity and performance requirements continue to increase, the probability of timing violations also increases. Even worse, even minor timing violations can lead to severe consequences such as functional failures or system crashes~\cite{gangadharan2013constraining}. As a result, efficient repair of timing violations has become an indispensable part of advanced VLSI implementation.

\begin{figure}[!t]
    \centering
    \includegraphics[width=0.45\textwidth]{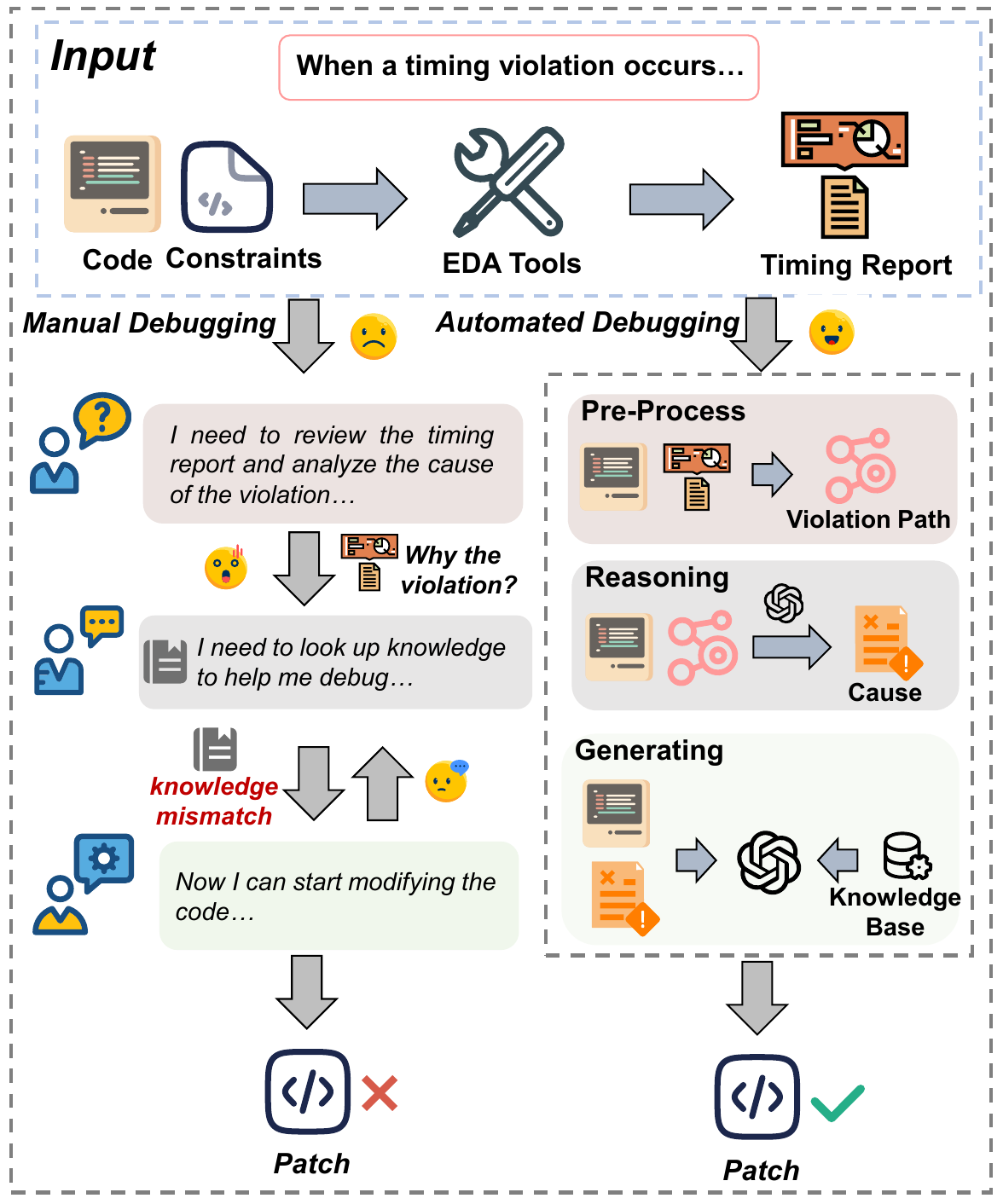}
    \caption{An example to illustrate the timing violation debug task.Formerly ``Manual Debugging'' workflow required manual intervention at every stage, which error-prone and inefficient.
    ViTAD integrates automation tools and LLMs for automated debugging to ensure less manual intervention and more efficient debugging.
}
    \label{fig:motivation}
    \vspace{-2em}
\end{figure}

\begin{figure*}[!ht]
    \centering
    \includegraphics[width=\textwidth, height=10.5cm]{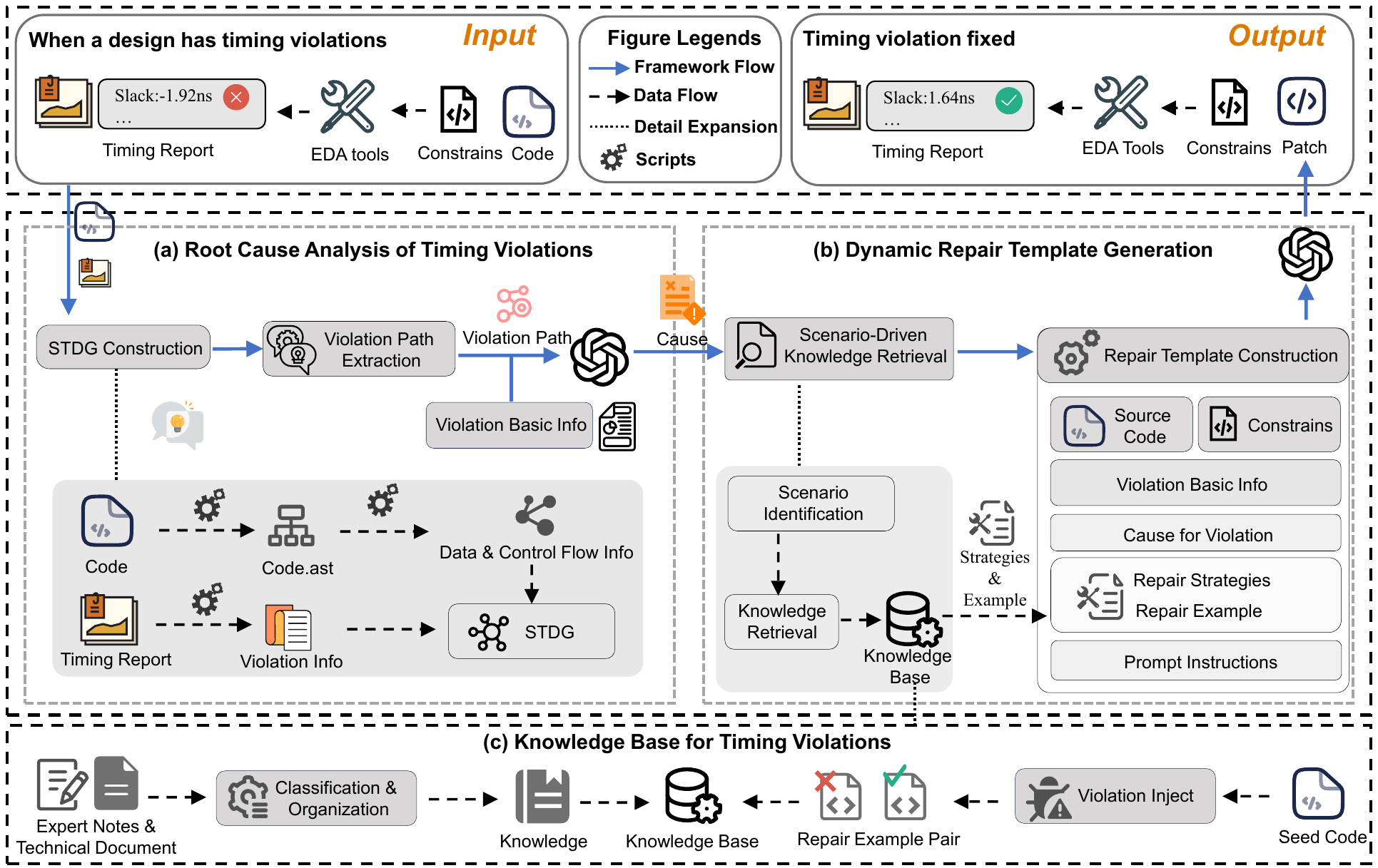}
    \caption{The overall pipeline of ViTAD. (a) The \textit{Root Cause Analysis of Timing Violations} module parses the code and reports to analyze the causes of violations; (b) The \textit{Dynamic Repair Template Generation} module identifies specific scenarios, retrieves repair strategies and automatically generates specific repair templates; (c) The \textit{Knowledge Base for Timing Violations} module collects and organizes debugging knowledge and repair demonstrations.
}
    \label{fig:overview}
    \vspace{-1em}
\end{figure*}

Nevertheless, traditional timing debugging methods remain a serious bottleneck in VLSI implementation. Specifically, these methods primarily combine Electronic Design Automation (EDA) timing verification tools~\cite{barbhaya2023open} with manual code inspection. Although EDA tools can accurately locate violations after the physical design stage, the process of root cause analysis and repair still depends heavily on engineers’ domain expertise. Consequently, as circuit complexity increases, even experienced engineers require more time to debug, often iterating through timing reports repeatedly until the issue is resolved.

Recently, Large Language Models (LLMs) have demonstrated remarkable performance in code generation and debugging tasks~\cite{yang2023asteria,abdollahpour2024automatic,valueian2022siturepair}. Despite this progress, most existing debugging approaches~\cite{tsai2024rtlfixer,xu2024meic} focus on syntax and functional issues, with limited attention to timing violations. Meanwhile, for timing optimization, many recent efforts remain at the prediction level, applying various techniques to estimate RTL timing behavior~\cite{fang2023masterrtl,sengupta2023early,lopera2021rtl}. In contrast, effective debugging methods for timing violations are still underexplored.

Figure \ref{fig:motivation} shows an example of the motivation for automatic timing violation debugging. In manual debugging workflows, engineers rely on timing reports generated by EDA tools to identify the root cause of violations. However, due to a lack of proper understanding of violations and reasoning tools, this process is prone to failure and often requires multiple attempts. Specifically, engineers interpret complex timing paths without automated assistance, leading to misdiagnosis of the root cause. Additionally, engineers with limited debugging experience may adopt incorrect repair solutions, resulting in failed repairs. Therefore, manual debugging is often inefficient and error-prone due to limited engineer experience, fragmented knowledge, and the lack of systematic analysis between code and timing behavior. To address this issue, integrating static analysis techniques with LLMs enables automated reasoning of violation causes and the generation of targeted code patches.

To automate the debugging of timing violations, we propose ViTAD, a novel framework that systematically integrates static analysis and LLM reasoning. Specifically, we first extract control flow and data flow information from Verilog code and integrate it with violation reports to construct a Signal Timing Dependency Graph (STDG). Leveraging the STDG, we perform violation path analysis to identify the root causes of timing violations using LLM. Based on the inferred causes, we selectively retrieve relevant debugging strategies from a domain-specific knowledge base to generate tailored repair solutions. By reducing manual intervention, ViTAD enables an automated and knowledge-guided approach to timing violation debugging.

Our main contributions are as follows:

\begin{itemize}[leftmargin=*, itemsep=0.05mm]
    \item We construct a dedicated knowledge base for timing violations. This knowledge base categorizes typical violation scenarios and organizes corresponding debugging knowledge to support timing violations debugging. 
    \item We propose a hybrid analysis method based on multi-dimensional feature fusion. By constructing a Signal Timing Dependency Graph (STDG), we enable joint modeling of code structure and timing information to assist LLMs in reasoning about violation causes.  
    \item We design a parameterized dynamic repair template engine. This engine analyzes specific violation causes and flexibly generates targeted repair strategies, enabling precise alignment between violation and repair solutions.  

\end{itemize}

\section{Preliminaries}

\subsection{Timing Violations and Debugging}

In digital circuits, data is expected to transfer correctly between registers on clock edges, requiring signals to arrive within defined timing constraints. A timing violation occurs when these constraints are not met—typically due to a signal arriving too late (setup violation) or changing too soon (hold violation), potentially causing incorrect computations or system instability. Additionally, Violations can also occur when signals cross between different clock domains, known as clock domain crossing (CDC) issues, which may lead to metastability or data loss.

In RTL designs, such as those written in Verilog, these constraints are determined by the logical and structural paths between registers. Timing violations are common in complex designs with deep pipelines, long combinational paths, or multiple clock domains. Debugging such violations goes beyond detection—it requires pinpointing the faulty paths, analyzing the root cause, and modifying the design to meet timing without altering its intended functionality.

\section{Methodology}

\subsection{Overview of ViTAD}
Figure \ref{fig:overview} shows an overview of ViTAD. ViTAD integrates an LLM for code repair, a structured timing violation knowledge base for guidance, and a graph-based root cause analysis engine. It first parses Verilog code and timing reports into a Signal Timing Dependency Graph (STDG), enabling precise violation path extraction. The root cause is then used to retrieve scenario-specific repair knowledge, which guides dynamic prompt construction. The LLM generates a targeted fix based on this context.

\subsection{Knowledge Base for Timing Violations}

We construct a domain-specific knowledge base for Verilog timing violations. By providing relevant domain information to the LLM, we aim to improve its debugging capabilities.

\begin{table}[!t]
    \fontsize{14}{14}
    \centering
    \renewcommand{\arraystretch}{1.8} 
    \label{tab:classification}
    \caption{Subscenario classification of violation types.}
    \resizebox{1.0\linewidth}{!}{
        \begin{threeparttable}
            \begin{tabular}{>{\centering\arraybackslash}m{3cm}|>{\centering\arraybackslash}m{4cm}|p{6cm}}
            \Xhline{0.8pt}
            \multicolumn{1}{c|}{\textbf{Types}} & \multicolumn{1}{c|}{\textbf{Subscenario}} & \multicolumn{1}{c}{\textbf{Description}} \\
            \Xhline{0.8pt}
            
            \multirow{3}{*}[0.8em]{\raisebox{-24ex}{\textbf{setup violation}}} & \LARGE \raisebox{-4ex}{long\_comb\_chain} & \Large Too many combinational logic stages or excessive delay causes data to arrive at the destination register later than required by clock cycle timing. \\ \cline{2-3}
            
                & \LARGE \raisebox{-3ex}{deep\_mux} & \Large Multi-level nested multiplexers create long selection delay paths, causing data propagation times to exceed clock cycle constraints. \\ \cline{2-3}
                & \LARGE \raisebox{-3ex}{low\_pipe\_stage} & \Large Complex operations cannot be completed within a single clock cycle but the current design lacks sufficient pipeline stages. \\ \cline{1-3}
        
            \raisebox{-5ex}{\textbf{hold violation}} & \LARGE \raisebox{-3.5ex}{short\_logic\_path} & \Large The combinational logic delay is too small, causing data to change before sampling is completed in the target register. \\ \cline{1-3}
            \multirow{2}{*}[0.3em]{\raisebox{-11ex}{\textbf{CDC error}}} & \LARGE \raisebox{-2ex}{single\_bit\_trans} & \Large Single-bit signals are transmitted directly between clock domains without proper synchronizers. \\ \cline{2-3}
            
                & \LARGE \raisebox{-3ex}{multi\_bits\_trans} & \Large Multi-bit data is transferred directly between clock domains without appropriate cross-domain transfer mechanisms. \\
            \Xhline{0.8pt}
            \end{tabular}
        \end{threeparttable}    
}
\end{table}

\subsubsection{Classification and Organization}

First, to extract expert knowledge, we obtain standardized descriptions and debugging rules from technical manuals and engineering notes. Second, we format the collected knowledge to ensure consistency and usability. Each knowledge entry adheres to a unified template comprising key fields—such as error descriptions and repair strategies—to ensure semantic clarity and completeness for retrieval. To support efficient organization and precise retrieval, we propose a three-level semantic classification scheme: \textit{Violation Type → Subscenario → Repair Strategy}. Each violation type is further categorized into sub-scenarios based on specific underlying mechanisms, as illustrated in Table 1. This hierarchical structure facilitates fine-grained access to domain knowledge.

\subsubsection{Violation Inject}
To help LLM effectively understand and simulate the debugging process, we constructed demonstration pairs consisting of fault code fragments and correction code fragments. These pairs are based on representative seed modules selected from open-source Verilog projects. Each module has been manually adapted to preserve critical timing logic. Additionally, we introduce timing violation into these modules using two complementary methods: (1) automatic injection, which simulates common timing errors such as setup violations, hold violations, and clock domain crossing (CDC) errors; (2) manual injection, which allows precise control over the nature and location of faults, particularly in complex or subtle scenarios.

\subsection{Root Cause Analysis of Timing Violations}

We propose a root cause analysis method for timing violations in Verilog code, inspired by the typical debugging workflow of tracing execution paths to locate error sources.

\subsubsection{Signal Timing Dependency Graph Construction}

\begin{figure}[!t]
    \centering
    \includegraphics[width=0.5\textwidth]{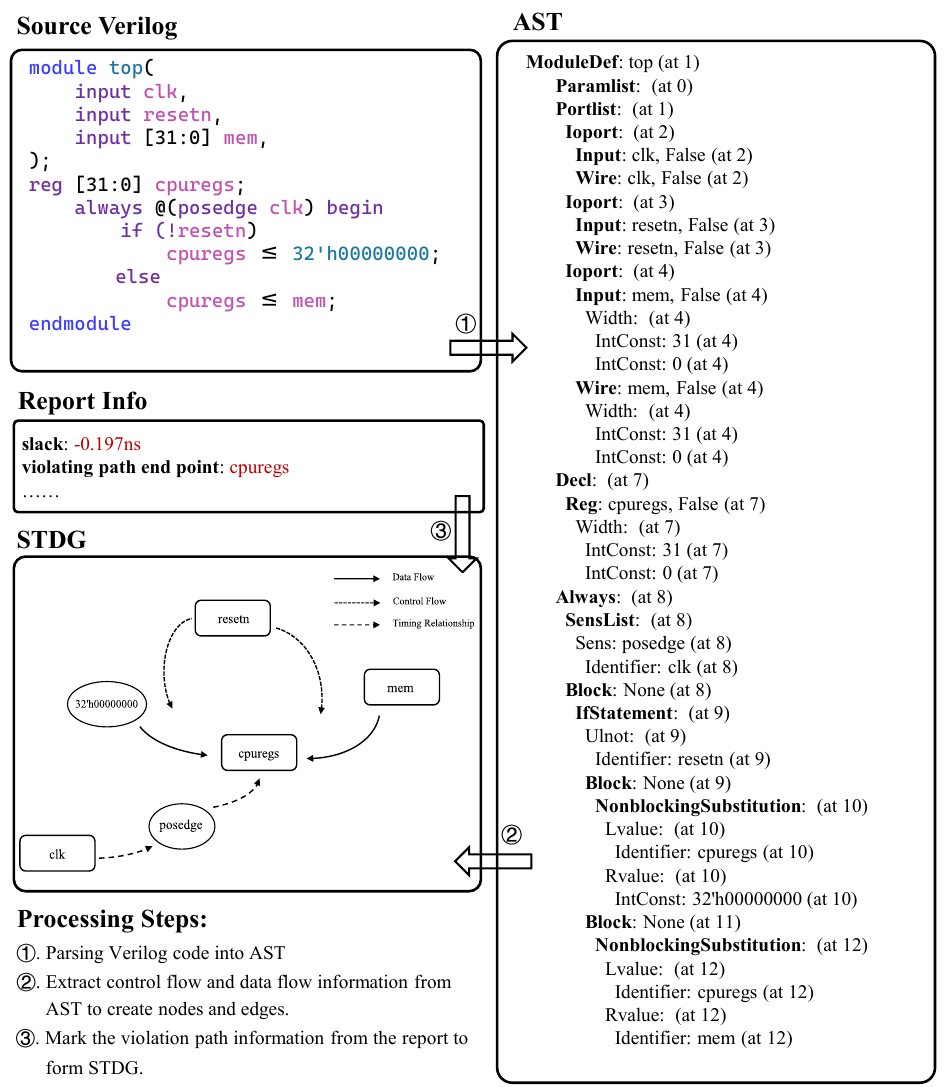}
    \caption{Example of creating a STDG from Verilog via AST Conversion and violation report.
}
    \label{fig:stdg_case}
    \vspace{-1em}
\end{figure}

We use tools~\cite{takamaeda2015pyverilog} to parse the Verilog code to Abstract Syntax Tree (AST) and extract control and data flow information, then integrate it with timing violation reports to build the Signal Timing Dependency Graph (STDG). Figure \ref{fig:stdg_case} illustrates this process. This graph enables full execution trace analysis of timing violation paths. In the STDG, nodes represent entities such as signals and registers in the Verilog design. Edges define relationships between these entities. Additionally, each node is annotated with information such as its source code location and bit-width. Register nodes involved in violations also carry additional timing violation information (from the timing report).  Edges in the graph describe both data and control dependencies between nodes.

\subsubsection{Violation Path Extraction}

After constructing the STDG by integrating structural code information with timing violations, we extract execution paths that correspond to the reported violations. This step is essential for linking violation reports to specific runtime behaviors and guiding accurate timing-aware debugging. Our path extraction algorithm starts from the violation register node in STDG and traverses the graph to identify all potential preceding nodes. Multiple logical paths may be generated in the initial stage because the target register may be accessible from multiple entry points. Subsequently, we simplify these paths into a single physical path, which represents the actual execution trajectory. The computation of the violation path can be formalized as:

\begin{equation}
\text{ViolationPath}(v_{\text{viol}}) = \scalebox{0.9}{$\mathcal{M} \left( \bigcup_{p \in \mathcal{T}(v_{\text{viol}}, G)} p \right)$}\,,
\end{equation}
where \( v_{\text{viol}} \) denotes the violation node, and \( \mathcal{T}(v_{\text{viol}}, G) \) represents the set of all logical paths originating from this node in the graph \( G \). The operator \( \bigcup \) aggregates all such paths, and \( \mathcal{M} \) maps them to the single path. \( \text{ViolationPath}(v_{\text{viol}}) \) is the final result.

Finally, the complete execution path is integrated with the violation code and basic violation information and fed into the LLM for violation cause analysis.

\begin{algorithm}[!t]
\caption{Scenario Identification from Violation Cause}
\label{alg:scenario_identification}
\begin{algorithmic}[1]
\Require Violation cause description $V$
\Ensure Identified violation scenario $S$

\State \textbf{Step 1:} Semantic Parsing
\State \quad Extract key parameters $\mathcal{P}$ from $V$ using regex and rules
\State \quad (e.g., signal names, delay values, path depth)

\State \textbf{Step 2:} Rule-Based Scenario Matching
\For{each predefined scenario $S_i \in \{S_1, S_2, ..., S_n\}$}
    \State Load scenario rule set $R_{S_i}$
    \If{$\mathcal{P}$ matches keywords and numerical patterns in $R_{S_i}$}
        \State \Return $S \gets S_i$
    \EndIf
\EndFor

\State \Return $S \gets$ \texttt{Unknown} \Comment{No matching scenario found}
\end{algorithmic}
\end{algorithm}

\subsection{Dynamic Repair Template Generation}

We design a dynamic template generation mechanism that adapts to different timing violation types and scenarios, enabling the LLM to produce accurate and context-aware repair prompts.

\subsubsection{Scenario-Driven Knowledge Retrieval} 
We designed a scenario identification-based method for retrieving knowledge. This method performs deep parsing of the technical semantics within the violation cause. It extracts key parameters and retrieves corresponding repair strategies and repair examples from a knowledge base based on the identified scenario. To effectively extract technical elements (circuit design information, timing constraints, error parameters) from the violation cause, we developed a rule-based multilevel parsing algorithm. Specifically, we construct dedicated rule sets for six scenarios of timing violations. Each set has two layers: (1) Regex patterns to identify technical terms and numerical parameters; (2) Scenario classification rules that match extracted parameter sets to violation scenarios. The process of identifying violation scenarios is elaborated in Algorithm 1.

\subsubsection{Repair Template Construction} Based on the retrieved knowledge, a mechanism dynamically builds repair templates for each violation type. It uses adaptive template structures and parameterized filling methods to generate personalized repair templates. Recognizing the distinct needs of setup/hold violations and CDC errors, we designed multiple structured templates. Intelligent parameter mapping ensures precise insertion of knowledge and other template parameters into the template structure, generating a complete guidance template.

\section{Experiments}

\begin{table*}[!ht]
    \small
    \setlength{\tabcolsep}{2pt} 
    \centering
    \label{tab:result}
    \renewcommand{\arraystretch}{1.8} 
    \caption{The timing violation debugging with different LLMs. The highest FR of each module is marked.}
    \resizebox{1.0\linewidth}{!}{
        \begin{threeparttable}
            \begin{tabular}{l|c|c|c|c|c|c|c|c|c|c}
            \Xhline{0.8pt}
            
            \multicolumn{2}{c|}{\textbf{Types}} & \multicolumn{1}{c|}{\textbf{GPT-4o}} & \multicolumn{1}{c|}{\makecell{\textbf{GPT-4o}\\\textbf{+Knowledge}}} & \multicolumn{1}{c|}{\makecell{\textbf{GPT-4o}\\\textbf{+ViTAD}}} & \multicolumn{1}{c|}{\textbf{Deepseek-R1}} & \multicolumn{1}{c|}{\makecell{\textbf{Deepseek-R1}\\\textbf{+Knowledge}}} & \multicolumn{1}{c|}{\makecell{\textbf{Deepseek-R1}\\\textbf{+ViTAD}}} & \multicolumn{1}{c|}{\textbf{Claude3.5}} & \multicolumn{1}{c|}{\makecell{\textbf{Claude3.5}\\\textbf{+Knowledge}}} & \multicolumn{1}{c}{\makecell{\textbf{Claude3.5}\\\textbf{+ViTAD}}} \\ 

            \Xhline{0.8pt}
            
            \multirow{2}{*}{\textbf{long\_comb\_chain}} & simple & 58.82\% & 47.05\% & 82.35\% & 70.58\% & 76.47\% & \cellcolor{green!30} 88.23\% & 52.94\% & 64.70\% & 76.47\% \\ \cline{2-11}
            
                & complex & 35.29\% & 52.94\% & 58.82\% & 41.17\% & 47.05\% & 52.94\% & 41.17\% & 47.05\% & \cellcolor{green!30} 58.82\% \\ \cline{1-11}
        
            \multirow{2}{*}{\textbf{deep\_mux}} & simple & 47.36\% & 52.63\% & 68.42\% & 57.89\% & 52.63\% & \cellcolor{green!30} 78.94\% & 42.10\% & 57.89\% & 63.15\% \\ \cline{2-11}
        		
        		&complex&63.15\%&47.36\%&52.63\%&52.63\%&42.10\%&63.15\%&57.89\%&63.15\%&\cellcolor{green!30}78.94\% \\ \cline{1-11}

            \multirow{2}{*}{\textbf{low\_pipe\_stage}} &simple&57.14\%&66.66\%&76.19\%&61.90\%&57.14\%&80.95\%&52.38\%&66.66\%&\cellcolor{green!30}85.71\% \\ \cline{2-11}
        		
        		&complex&52.38\%&57.14\%&\cellcolor{green!30}66.66\%&47.16\%&52.38\%&57.14\%&42.85\%&47.61\%&61.90\% \\ \cline{1-11}

            \multirow{2}{*}{\textbf{short\_logic\_path}} &simple&58.82\%&64.70\%&76.47\%&64.70\%&76.47\%&\cellcolor{green!30}88.23\%&58.82\%&47.05\%&70.58\% \\ \cline{2-11}
        		
        		&complex&52.94\%&47.05\%&64.70\%&58.82\%&64.70\%&\cellcolor{green!30}70.58\%&64.70\%&35.29\%&47.05\% \\ \cline{1-11}

            \multirow{2}{*}{\textbf{single\_bit\_trans}} &simple&78.94\%&84.21\%&\cellcolor{green!30}100\%&84.21\%&89.47\%&\cellcolor{green!30}100\%&89.47\%&94.73\%&\cellcolor{green!30}100\% \\ \cline{2-11}
        		
        		&complex&36.84\%&42.10\%&63.15\%&47.36\%&42.10\%&\cellcolor{green!30}68.42\%&42.10\%&47.36\%&52.63\% \\ \cline{1-11}

            \multirow{2}{*}{\textbf{multi\_bits\_trans}} &simple&61.90\%&66.66\%&76.19\%&71.42\%&76.19\%&\cellcolor{green!30}85.71\%&66.66\%&71.42\%&76.19\% \\ \cline{2-11}
        		
        		&complex&47.61\%&57.14\%&\cellcolor{green!30}66.66\%&42.85\%&52.38\%&57.14\%&52.38\%&57.14\%&61.90\% \\ \cline{1-11}

            \multicolumn{2}{c|}{\fontsize{10}{10}\textbf{FR}} & 54.38\% & 57.45\% & 71.05\% & 58.33\% & 60.52\% & \cellcolor{green!30} 73.68\% & 55.26\% & 58.77\% & 69.73\%  \\ 
                
            \Xhline{0.8pt}
            \end{tabular}
        \end{threeparttable}    
}

\end{table*}

\subsubsection{Setup}In our experiments, GPT-4o~\cite{achiam2023gpt}, Deepseek-R1~\cite{guo2025deepseek}, and Claude3.5 were used as LLMs for all experiments, except for GPT-4o in the ablation experiment. We set the temperature of LLM to 0.7 to control the randomness of LLM output. In terms of hardware, we used the Vivado tool, with the device selected as the XC7A100TCSG324-1 for simulating physical layout routing and timing analysis.  To limit the test variance, the debugging cases in each experiment are repeated multiple times.

\subsubsection{Dataset}To facilitate the evaluation of our proposed timing violation analysis and repair methods, we constructed a high-quality Verilog dataset tailored for timing violations. We selected RTL designs with clear structural and modular boundaries from several mainstream open-source projects (e.g., PicoRV32, OpenRISC) as seed designs, into which we systematically injected timing violations. Additionally, we further categorize the faulty designs into two levels—simple and complex—based on their logical complexity and the severity of the timing violation. It is worth noting that simple does not imply small code size, but rather refers to the relative ease of debugging compared to complex cases. In summary, the dataset encompasses a range of realistic and challenging timing violations, including setup violations, hold violations, and CDC errors, totaling 54 cases.

\subsubsection{Baselines}

We evaluate ViTAD against two groups of baseline models, each consisting of three different LLMs: GPT-4o, Deepseek-R1, and Claude 3.5. The first group comprises standard models, which attempt repairs by directly receiving the faulty code along with basic contextual information. The second group consists of enhanced standard models that are supplemented with additional domain-specific knowledge. Unlike the first group, these models receive extra domain-specific inputs as part of the prompt context.

\subsubsection{Evaluation Metric}

Fix rate: To demonstrate the debugging ability of our method, we calculated the expected repair rate, where c is the number of successfully repaired samples among all n samples.

\begin{equation}
    \text{fix rate} = \mathbb{E}_{\text{problems}} \left[ \frac{c}{n} \right]
    \label{eq:metric}
\end{equation}

\subsubsection{Research Questions}

Our work aims to answer four research questions:

\begin{itemize}
    \item \textbf{RQ1:}How does various LLM-based configurations and integration impact debugging performance in terms of fix rate?
    \item \textbf{RQ2:}Can ViTAD correct different types of errors in more complex scenarios? 
    \item \textbf{RQ3:}How does ViTAD work with different LLM?  
    \item \textbf{RQ4:}What are the components that affect ViTAD debugging results?
\end{itemize}
We also present a case study of an unsuccessful repair attempt to identify key factors contributing to the failure, which may guide future improvements.

\subsection{Effectiveness of ViTAD (RQ1)}

This RQ investigates whether the proposed ViTAD framework can significantly improve the debugging success rate for timing violations. Table 2 reports the Fix Rate (FR) achieved by three LLMs (GPT-4o, Deepseek-R1, and Claude 3.5) under three configurations: baseline (standard LLMs), LLMs augmented with explicit domain knowledge, and LLMs integrated with ViTAD.The results show a consistent improvement in FR when ViTAD is applied. For instance, GPT-4o, Deepseek-R1, and Claude3.5 reached 71.05\%, 73.68\%, and 69.73\% FR with ViTAD, compared to 54.38\%, 58.33\%, and 55.26\% with only domain knowledge. All models in their baseline forms achieved the lowest FR. These findings suggest that while adding domain-specific knowledge improves performance over the baseline, ViTAD provides a more substantial boost. Interestingly, the gains are especially significant in complex timing violation scenarios, where structured reasoning and scenario-specific strategies within ViTAD are more effective. This demonstrates that ViTAD not only enhances standard LLMs but also outperforms models that rely solely on external knowledge.

\begin{figure}[!t]
    \centering
    \begin{subfigure}{0.45\textwidth}
        \includegraphics[width=\linewidth]{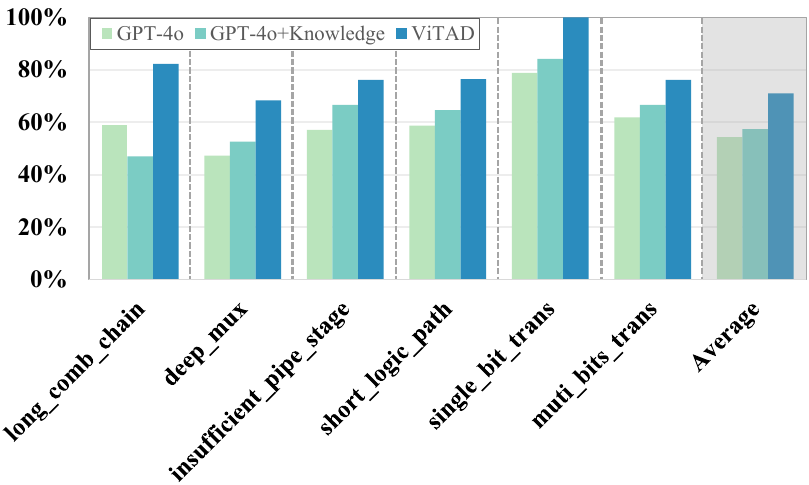}
        \caption{FR of simple timing violations.}
        \label{fig:subfig1}
    \end{subfigure}
    \hfill
    \begin{subfigure}{0.45\textwidth}
        \includegraphics[width=\linewidth]{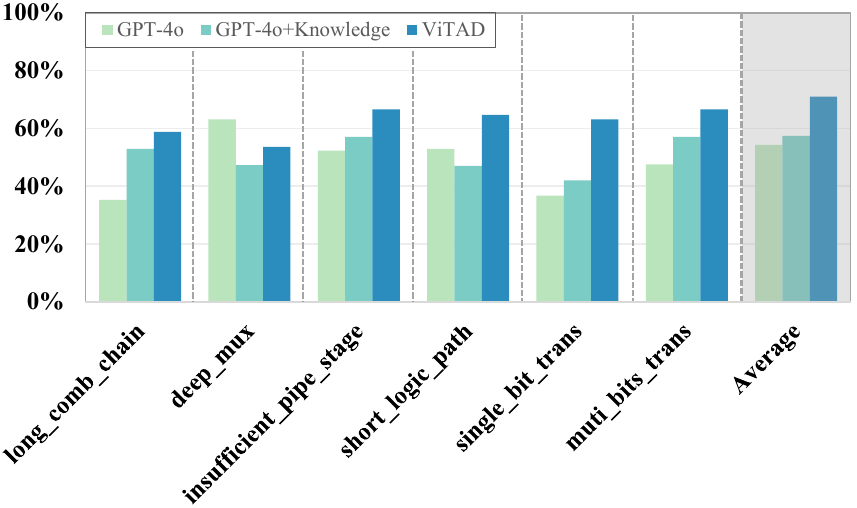}
        \caption{FR of complex timing violations.}
        \label{fig:subfig2}
    \end{subfigure}
    \caption{FR in timing violations with different Method.}
    \label{fig:main}
    \vspace{-1em}
\end{figure}

\subsection{Generalizability Across Scenarios (RQ2)}

This RQ aims to evaluate the effectiveness of the ViTAD framework in resolving timing violations under a range of scenarios. It seeks to understand how the system's debugging performance varies with the complexity of the code and the specific type of violation. Figure \ref{fig:main} presents the Fix Rate (FR), as defined in Equation \ref{eq:metric}, achieved by ViTAD across six common timing violation scenarios. The results reveal significant variation in FR, which correlates with both the scenario types and their inherent complexity. For example, in simpler scenarios such as the "short\_logic\_path" with basic combinational logic, ViTAD achieves consistently high FRs, showcasing its strong capability in addressing mild violations. In contrast, more complex cases like "multi\_bits\_trans" show a noticeable drop in FR, underscoring the challenges posed by intricate timing faults. Overall, ViTAD demonstrates high effectiveness on simpler violations and maintains solid performance in more complex cases, supporting its practical value in realistic timing debugging tasks.

\subsection{LLM Integration Impact (RQ3)}

This RQ investigates the comparative effectiveness of integrating different LLMs—specifically GPT-4o, Deepseek-R1, and Claude3.5—into ViTAD for debugging code. It aims to understand how these models influence the performance of ViTAD by quantifying their effectiveness in timing violation debugging. As shown in Table 2, the standard Deepseek-R1 baseline outperforms GPT-4o across most scenarios. Even when augmented with external knowledge, GPT-4o fails to match the performance of the standard Deepseek-R1. This result highlights the differences in the models’ inherent capabilities when handling long and complex timing violation inputs. However, when integrated into ViTAD, GPT-4o achieves an FR of 71.05\%, surpassing the standard Deepseek-R1—indicating that our framework effectively enhances the model’s debugging performance. This further confirms the effectiveness of our framework in enhancing LLM-guided debugging for timing violations.

\subsection{Key Design Contributions (RQ4)}

This RQ investigates the contributions of ViTAD’s key components. Specifically, we conduct two ablation experiments to assess the impact of violation cause descriptions and repair guidance on timing violation repair. We evaluate four settings: Set 1 (w/o cause \& guidance): Neither input is provided. Set 2 (w/o cause): Only repair guidance is given. Set 3 (w/o guidance): Only cause descriptions are provided. Set 4 (ViTAD): Both are included.

As shown in Table 3, the worst performance occurs in Set 1. Set 2 shows minor improvement, while Set 3 performs better, highlighting the value of causality. Set 4 achieves the highest fix rate (FR), confirming the effectiveness of combining both inputs. These results show that cause descriptions, especially explicit violation paths, greatly improve repair success by helping LLMs locate root issues. The small gap between Set3 and Set4 suggests that clear causality enables LLMs to infer repair strategies even without explicit guidance. Overall, both components are beneficial, with causality being the key driver.

\begin{table}[!t]
    \small
    \centering
    \label{tab:ablation}
    \renewcommand{\arraystretch}{1.8} 
    \caption{Ablation study on ViTAD.}
    \resizebox{1.0\linewidth}{!}{
        \begin{threeparttable}
            \begin{tabular}{c|c|c|c|c|c}
            \Xhline{0.8pt}
            \multicolumn{2}{c|}{\textbf{Types}} & \multicolumn{1}{c|}{\makecell{\textbf{w/o }\\\textbf{cause\&guidance}}} & \multicolumn{1}{c|}{\makecell{\textbf{w/o}\\\textbf{cause}}} & \multicolumn{1}{c|}{\makecell{\textbf{w/o}\\\textbf{guidance}}} & \multicolumn{1}{c}{\textbf{ViTAD}} \\ \cline{1-6}
            
            \multirow{2}{*}{\textbf{long\_comb\_chain}} &simple&58.82\%&64.70\%&76.47\%&\cellcolor{green!30}82.35\% \\ \cline{2-6}
            
                &complex&35.29\%&41.17\%&52.94\%&\cellcolor{green!30}58.82\% \\ \cline{1-6}
        
            \multirow{2}{*}{\textbf{deep\_mux}} &simple&47.36\%&52.63\%&57.89\%&\cellcolor{green!30}68.42\% \\ \cline{2-6}
            
                &complex&\cellcolor{green!30}63.15\%&57.89\%&52.63\%&52.63\% \\ \cline{1-6}

            \multirow{2}{*}{\textbf{low\_pipe\_stage}} &simple&57.14\%&71.42\%&71.42\%&\cellcolor{green!30}76.19\% \\ \cline{2-6}
            
                &complex&52.38\%&42.85\%&61.90\%&\cellcolor{green!30}66.66\% \\ \cline{1-6}

            \multirow{2}{*}{\textbf{short\_logic\_path}} &simple&58.82\%&64.70\%&70.58\%&\cellcolor{green!30}76.47\% \\ \cline{2-6}
            
                &complex&52.94\%&58.82\%&58.82\%&\cellcolor{green!30}64.70\% \\ \cline{1-6}

            \multirow{2}{*}{\textbf{single\_bit\_trans}} &simple&78.94\%&89.47\%&100\%&\cellcolor{green!30}100\% \\ \cline{2-6}
            
                &complex&36.84\%&47.36\%&52.63\%&\cellcolor{green!30}63.15\% \\ \cline{1-6}

            \multirow{2}{*}{\textbf{multi\_bits\_trans}} &simple&61.90\%&66.66\%&71.42\%&\cellcolor{green!30}76.19\% \\ \cline{2-6}
            
                &complex&47.61\%&57.17\%&61.90\%&\cellcolor{green!30}66.66\% \\ \cline{1-6}

            \multicolumn{2}{c|}{\textbf{FR}} & 54.38\% & 59.64\% & 65.78\% & \cellcolor{green!30}71.05\%   \\ \cline{1-6}

            \end{tabular}
        \end{threeparttable}    
}

\end{table}

\subsection{Case Study}
\begin{figure}[!t]
    \centering
    \includegraphics[width=0.48\textwidth]{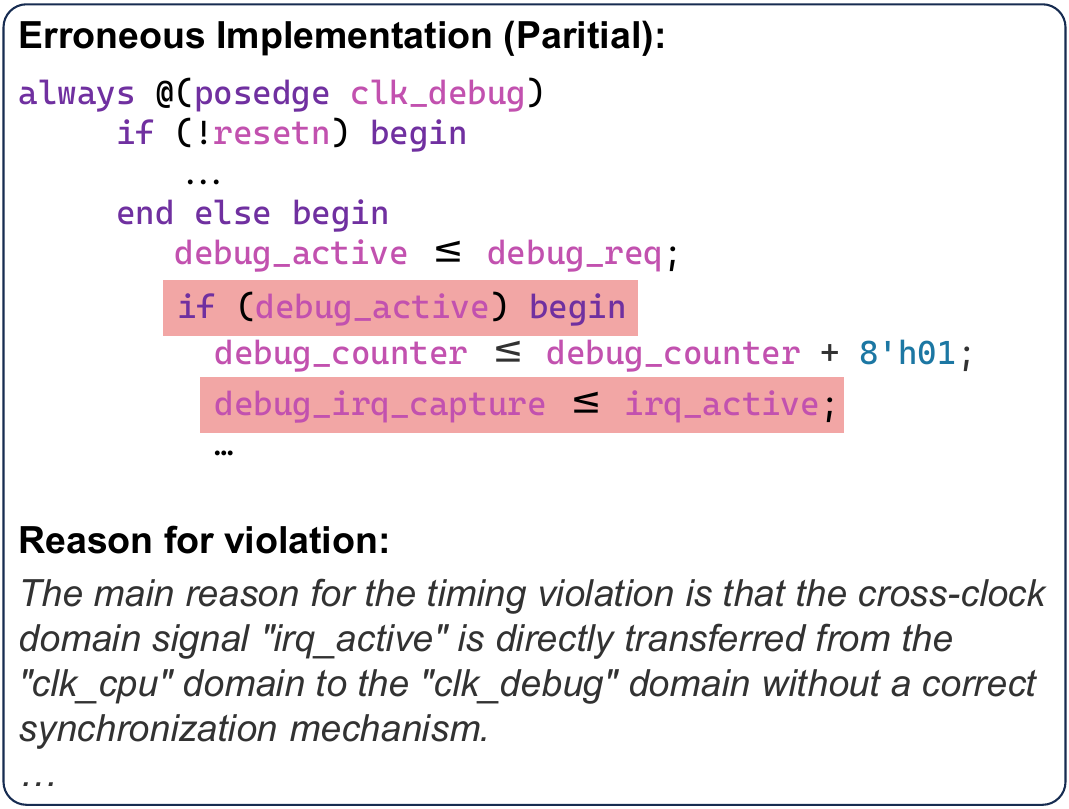}
    \caption{An example which the ViTAD failed to fix a timing violation. LLM cannot accurately determine that the control signal cannot be used to control asynchronous transmission
}
    \label{fig:bad example}
    \vspace{-1em}
\end{figure}

In this study, we adopt an LLM-based framework to repair timing violations and achieve a satisfactory overall success rate. However, reliance on LLMs presents challenges. In some scenarios, even when clear violation paths and relevant domain knowledge are provided, the LLM fails to identify the root cause correctly, potentially resulting in incorrect repairs. Figure \ref{fig:bad example} illustrates one such failure case.

A comparative analysis with successful repair cases reveals two possible reasons for this failure: (1) the code is relatively large and contains complex logic, which may exceed the LLM’s capacity to process long and intricate inputs; and (2) the violation may stem from multiple potential causes, while the provided prompts lack sufficient specificity for the LLM to identify the correct one in context. To address these issues, we believe that incorporating more precise domain-specific knowledge and introducing iterative or backtracking mechanisms may improve the handling of failed repair attempts, thereby enhancing the adaptability and robustness of the repair framework.

\section{Related Work}

\paragraph{LLM for General Code Debugging}

Recently, LLMs have demonstrated impressive performance in code-related tasks such as generation, editing, and repair. Existing approaches for applying LLMs to code repair can be broadly divided into two categories. The first category focuses on fine-tuning LLMs with large-scale code datasets to enhance repair effectiveness. For instance, Zhou et al. proposed VulMaster~\cite{zhou2024out}, which adopts a Fusion-in-Decoder architecture to incorporate full source code, AST structures, and CWE knowledge. Liu et al.~\cite{liu2024fastfixer} introduced a method that combines modification-oriented fine-tuning with parallel validation. Huang et al.~\cite{huang2025template} developed a two-stage fine-tuning scheme that first selects a repair template, then generates code accordingly—effectively injecting traditional repair patterns into LLMs. The second category enhances prompt quality using static analysis or domain knowledge, guiding LLMs without additional training. Nashid et al. \cite{nashid2023retrieval} presented a retrieval-based few-shot prompting technique, enabling LLMs to perform assertion or code generation directly. TypeFix~\cite{peng2024domain} designed a prompt-learning method that integrates mined general templates into repair prompts. Ouyang et al.~\cite{ouyang2025knowledge} proposed a knowledge-enhanced framework for repairing data science code by retrieving API information and augmenting error descriptions via knowledge graphs. Ruan et al.~\cite{ruan2024specrover} leveraged LLMs for code intent extraction and applied an iterative specification inference and feedback mechanism to boost repair quality.

\paragraph{LLM for Hardware Code Debugging}

Traditional approaches rely on STA and formal verification to detect timing violations and functional bugs, but fixing them still requires manual effort. To address the inefficiency of traditional debugging, LLMs have been applied to automate hardware code repair. Tsai et al. \cite{tsai2024rtlfixer} combined LLMs with reactive frameworks, using iterative "think-execute-feedback" loops to fix syntax errors. Xu et al. \cite{xu2024meic} integrated multi-agent LLMs with RTL toolchains to perform pipelined debugging of functional errors via version-controlled rollbacks. Ahmad et al. \cite{ahmad2024hardware} further extended LLMs to patch security vulnerabilities. These methods mainly target syntax and functional issues, with limited focus on timing violations. In contrast, our work focuses on debugging timing violations.

\paragraph{Specialized Timing Violation Debugging}
Recent studies have explored the application of LLMs to timing-specific tasks, although primarily for prediction or analysis. RTL-Timer \cite{fang2024annotating} predicts register arrival times using ML-friendly RTL representations, but does not provide repair solutions. Amik et al. \cite{amik2025graph} convert RTL into graph structures for early timing estimation, but the representation lacks traceability for violations. The Timing Analysis Agent \cite{nainani2025timing} uses multi-agent LLMs to extract redundant metrics from STA reports but cannot generate code fixes.
In summary, existing work is not directly applicable to debugging timing violations. Most efforts focus on syntax or functionality, or remain limited to timing violation prediction and analysis. In practice, timing debugging is far more complex, requiring both domain expertise and a deeper understanding of violation causes.

\section{Conclusion}

ViTAD presents a novel LLM-integrated framework for automated timing violation debugging in RTL design. By unifying code structure and timing violation information into a signal timing dependence graph (STDG), our approach enables precise root cause analysis. The integration of a domain-specific knowledge base (categorized by violation scenarios) and a parameterized template engine enables dynamic generation of repair strategies. ViTAD achieves a success rate of 73. 68\% in debugging setup / hold violations and CDC errors through targeted RTL modifications. This demonstrates the feasibility of using LLMs to transform manual timing debugging into an automated, knowledge-driven workflow, reducing design iteration cycles.

\bibliography{test}

\end{document}